\documentclass[12pt,epsfig]{article}

\usepackage{epsfig}

\makeatletter
\renewcommand{\arraystretch}{1.5}

\newlength{\abstwidth}
\begin{document}

\begin{center}

\renewcommand{\arraystretch}{1.1}

{\LARGE \bf GaGaRes: A Monte Carlo generator for \\[3mm]
resonance production in two-photon physics} \\[10mm]

{\Large F.\ A.\ Berends$^a$ and R.\ van Gulik$^{a,b}$} \\[2ex]

{\it   
${}^a$Instituut-Lorentz, 
University of Leiden, P.O. Box 9506, 2300 RA Leiden,  
The Netherlands\\[2ex]
${}^b$NIKHEF, P.O. Box 41882, 1009 DB Amsterdam, The Netherlands
}

\end{center}

 
\begin{center}
{\bf Abstract}\\[2ex]
\begin{minipage}{\abstwidth}
In this paper a Monte Carlo generator, GaGaRes, is presented which
can be used to describe two-photon resonance production in $e^+e^-$
collisions. The program can generate the five lowest-lying ${\cal C}$=+1
meson states of any $q\bar{q}$ combination
together with the outgoing electron and positron.
The dependence on the photon virtualities
$Q_1^2$ and $Q_2^2$ is fully taken into account. The program 
also generates the density matrices of the resonance, which form an 
essential tool in the description of the decay  
of the resonances. Furthermore, the program is applicable for all tagging
conditions.
\end{minipage}
\end{center}


\vspace{1.0cm}

\noindent
{\bf {\Large PROGRAM SUMMARY}}

\noindent
{\it Title of program:} GaGaRes \\
{\it Program obtainable from:} 
CPC Library \\
{\it Computer for which the program is designed and 
others on which it is operable:} Any computer that runs under UNIX and
 can handle quadruple precision numbers (both real and complex). FORTRAN 90
offers the possibility to define own data types. This should be used
on systems where complex numbers are not available in quadruple 
precision. \\
{\it Operating system under which the program has been tested :} UNIX \\
{\it Programming language used :} FORTRAN 90 \\
{\it Number of lines:} 6009 (including comments) \\
{\it Keywords:} Monte Carlo, two-photon, $e^+e^-$, resonance production \\
{\it Nature of physical problem:} With the advent of LEP2 higher energies
for two-photon reactions became available with high luminosities. This makes
it possible
to search experimentally 
for heavier resonances created in two-photon collisions and 
also to determine 
the dependence of the two-photon cross sections on the virtualities $Q_1^2$
and $Q_2^2$. 
Moreover, the decay distributions of the resonances can be studied. 
These experimental possibilities make it desirable to have a program, which can
simulate events as expected from our theoretical understanding of resonance
production by two photons.
\\
{\it Methods of solution:} A model based on the hard scattering approach
 is used to describe
 the production of the resonances\cite{SchuBerGul}. 
For an exact description of the decay of the produced resonance
the density matrix is required. Weyl-van-der-Waerden spinor calculations
are used to obtain these density matrices. Events consisting of the momenta
of the resonance and the outgoing electron and positron are generated
by Monte Carlo methods and are distributed 
according to the theoretical cross section.\\
{\it Typical running time:} Depends on the requested accuracy and the
 generated resonance. On an  Enterprise450, one can produce
380 $^1S_0$ resonances/sec and 17 $^3P_2$ resonances/sec, 
including
the corresponding density matrix. Without density matrices, one can 
generate 236 $^3P_2$ resonances/sec. These numbers are for weighted events.
For unweighted events these rates are typically an order of magnitude lower.\\


\section{Introduction}
With the advent of LEP2 a new region for two-photon physics was entered. Its
high energy and its high luminosity, resulting in better statistics, allows
one to look for the dependence of the two-photon cross sections on the 
$\gamma\gamma$ centre-of-mass energy $W$ and on the photon virtualities
$Q_1^2$ and $Q_2^2$. The higher statistics also gives one the opportunity to
look for the decay distributions of the produced resonances.

The theoretical framework \cite{SchuBerGul} on which the program 
is based is obtained from the hard scattering approach in which 
the meson resonance is treated as a non-relativistic bound state of a quark
($q$) and an anti-quark ($\bar{q}$).
In order to include
also the production of lighter resonances some modifications had to be
implemented. These modifications are discussed in \cite{SchuBerGul}. The 
two-photon widths of the resonances that are calculated using this model 
give a reasonably good agreement with experiment, also for the lighter 
resonances.

The matrix elements following from this model have been implemented in 
the program, including all dependences on $W$ and $Q_i^2$. This is in contrast
to most other two-photon MC generators \cite{Linde,Buijs} where the $Q_i^2$
dependence is ad-hoc parametrized by 
multiplying the cross section by the square of a 
form factor $F^2(Q_1^2,Q_2^2)$.
The VMD model predicts the  pole-mass form factor 
\begin{equation}
  F(Q_1^2,Q_2^2)=\frac{M_V^2}{(M_V^2+Q_1^2)}
  \frac{M_V^2}{(M_V^2+Q_2^2)}.
\end{equation}
In this form factor the scaling mass $M_V$ is the mass of a resonance.
For the lighter resonances one takes $V=\rho$, for the $c\bar{c}$ resonances
$V=J/\psi$ whereas for the $b\bar{b}$ resonances $V=\Upsilon$.
Note that this form factor causes a very strong decrease of the cross section
with high $Q_i^2$'s. In \cite{SchuBerGul} it is discussed that the model 
used in GaGaRes results in a much weaker decrease of the cross section 
with high $Q_i^2$.

The total angular momentum of the resonance is denoted by $J$, its spin by
$S(=0,1)$ and its orbital angular momentum by 
$L(=0,1,2,\ldots \mbox{ or } S,P,D,\ldots)$, resulting in the well-known 
spectroscopic notation $^{(2S+1)}L_J$, with parity $P=(-1)^{L+1}$ and 
charge conjugation $C=(-1)^{L+S}$.

In two-photon reactions one can, in lowest order, only produce $C$-even
resonances. The present version of GaGaRes describes the production of 
$^1S_0$ $(0^-)$, $^3P_J \hspace{0.3cm} (J=0,1,2)$ $(J^+)$ 
and $^1D_2$ $(2^-)$ resonances. The resonance can be composed of 
light ($u$, $d$, $s$) quarks in the states $I=0,1$, where $I$ is the
isospin,  
or of heavy $c$ and $b$ quarks.

The outline of the paper is as follows. We start by introducing the 
kinematics and notation and present the different calculational methods  
to obtain the expressions for the matrix elements squared. Then the
discussion focusses on the density
matrices, which are required to describe the decay of the resonances. 
This is followed by a description of the two Monte Carlo  
generation
schemes available in GaGaRes for the integration of the
differential cross section. Finally, the structure of the program is presented,
together with some important variables, common blocks and procedures.


\section{Notation, Kinematics and Calculations of the Matrix Elements Squared}
The process of resonance production in two-photon physics is described by
the following reaction
\renewcommand{\arraystretch}{1.2}
\begin{equation}
  \begin{array}{l}
    e^+(p_1) + e^-(p_2) \to e^+(p_3) +e^-(p_4) + \gamma^*(k_1) +\gamma^*(k_2)
    \\ \hspace{0.5cm}
    \to  e^+(p_3) +e^-(p_4) + R(P).
  \end{array}
  \label{ReactionFormula}
\end{equation}
\renewcommand{\arraystretch}{1.0}

In this reaction the $\gamma^*$'s are  
off-shell photons which collide to
form the resonance. Four-momentum conservation in the vertices gives
\begin{equation}
  \begin{array}{lcl}
    k_1 & = & p_1 - p_3, \\
    k_2 & = & p_2 - p_4, \\
    P   & = & k_1 + k_2. 
  \end{array}
\end{equation}

The following parametrization of the external four-momenta in the lab-frame 
has been chosen.
\begin{equation}
  \begin{array}{lcl}
    p_1 & = & \left( E_b,0,0,-P_b \right), \\
    p_2 & = & \left( E_b,0,0,P_b \right),  \\
    p_3 & = & \left( E_3,|\vec{p}_3| \sin \theta_3 \cos \phi_3,
    |\vec{p}_3| \sin \theta_3 \sin \phi_3, -|\vec{p}_3| \cos\theta_3 \right),\\
    p_4 & = & \left( E_4,|\vec{p}_4| \sin \theta_4 \cos \phi_4,
    |\vec{p}_4| \sin \theta_4 \sin \phi_4, |\vec{p}_4| \cos \theta_4 \right),\\
    P   & = & \left( E_R,|\vec{p}_R| \sin \theta_R \cos \phi_R,
    |\vec{p}_R| \sin \theta_R \sin \phi_R, |\vec{p}_R| \cos \theta_R \right).
  \end{array}
  \label{MomPar}
\end{equation}
In these formulae $E_b$ is the beam-energy. $P_b$ is the momentum of the 
incoming leptons.
The polar angles $\theta_3$ and $\theta_4$ are defined as the angle 
between the
incoming lepton and the corresponding outgoing lepton.

It is convenient to introduce the set of invariants
\renewcommand{\arraystretch}{1.1}
\begin{equation}
  \begin{array}{lcl}
    s  & = & (p_1 + p_2)^2, \\
    s' & = & (p_3 + p_4)^2, \\
    t  & = & k_1^2 = -Q_1^2 = (p_1-p_3)^2, \\
    t' & = & k_2^2 = -Q_2^2 = (p_2-p_4)^2, \\
    u  & = & (p_1-p_4)^2, \\
    u' & = & (p_2-p_3)^2, \\
    W^2 & = & P^2 = (k_1 + k_2)^2,
  \end{array}
  \label{Invariants}
\end{equation}
where $Q_1^2$ and $Q_2^2$ are called the virtualities of the intermediate
photons. These invariants satisfy the relation
\begin{equation} 
  s + s' + t + t' + u + u' = W^2 + 8m_e^2.
\end{equation}

Using the methods of \cite{SchuBerGul} expressions for
the matrix elements of reaction  
(\ref{ReactionFormula}) can be 
obtained\footnote{
%
For the Minkowski metric we use $(1,-1,-1,-1)$. The totally
antisymmetric Levi-Civita tensor is 
defined by
$\varepsilon_{0123}=+1$. Also the notation 
$\varepsilon[p,q,r,s] \equiv 
\varepsilon^{\mu\nu\lambda\rho}p_\mu q_\nu r_\lambda s_\rho$ is introduced.
}
\begin{equation}
  \begin{array}{lcl}
    {\cal M}(^1S_0;\lambda_1,\lambda_2,\lambda_3,\lambda_4) & = & 
    \frac{c_1}{tt'} \varepsilon \left[ k_1,k_2,j_1(\lambda_1,\lambda_3),
    j_2(\lambda_2,\lambda_4)\right], \\
    {\cal M}(^3P_0;\lambda_1,\lambda_2,\lambda_3,\lambda_4) & = &     
    \frac{c_2}{tt'} \left( \left[ j_1(\lambda_1,\lambda_3) \cdot 
    j_2(\lambda_2,\lambda_4) 
    k_1 \cdot k_2  \right. \right. \\
    && \left. - j_1(\lambda_1,\lambda_3) 
    \cdot k_2 j_2(\lambda_2,\lambda_4)
    \cdot k_1 \right] (W^2+ k_1 \cdot k_2) \\
    && \left.  - 
    j_1(\lambda_1,\lambda_3)\cdot j_2(\lambda_2,\lambda_4) t t' \right), \\
    {\cal M}(^3P_1;\lambda_1,\lambda_2,\lambda_3,\lambda_4,\lambda_R) & = & 
    \frac{c_3}{tt'} \left( t\varepsilon \left[ \varepsilon^*(\lambda_R),
    j_1(\lambda_1,\lambda_3), j_2(\lambda_2,\lambda_4) ,k_2 \right] \right. \\
    && \left. +
    t'\varepsilon \left[ \varepsilon^*(\lambda_R),
    j_2(\lambda_2,\lambda_4),j_1(\lambda_1,\lambda_3) ,k_1 \right] \right), \\
    {\cal M}(^3P_2;\lambda_1,\lambda_2,\lambda_3,\lambda_4,\lambda_R) & = & 
    \frac{c_4}{tt'} \left( k_1 \cdot k_2 j_{1\mu}(\lambda_1,\lambda_3)
    j_{2\nu}(\lambda_2,\lambda_4) \right. \\
    && + k_{1\mu}k_{2\nu}  j_1(\lambda_1,\lambda_3) \cdot 
    j_2(\lambda_2,\lambda_4)  \\
    && - k_{1\mu} j_{2\nu}(\lambda_2,\lambda_4)
    j_1(\lambda_1,\lambda_3) \cdot k_2 \\
    && \left. - k_{2\mu} j_{1\nu}(\lambda_1,\lambda_3)
    j_2(\lambda_2,\lambda_4) \cdot k_1 \right) 
    \varepsilon^{*\mu\nu}(\lambda_R),  \\
    {\cal M}(^1D_2;\lambda_1,\lambda_2,\lambda_3,\lambda_4,\lambda_R) & = & 
    \frac{c_5}{tt'}\varepsilon^{*\mu\nu}(\lambda_R)
    k_{1\mu} k_{2\nu} \\ &&
    \varepsilon \left[ k_1,k_2,j_1(\lambda_1,\lambda_3),
    j_2(\lambda_2,\lambda_4) \right].    
  \end{array}
  \label{StandardAmplitudes}
\end{equation}
In these formulae $\lambda_i$ represents the helicity of the associated
external
particle. The lepton currents $j_1$ and $j_2$ are defined by
\begin{equation}
  \begin{array}{lcl}
    j_1^\mu(\lambda_1,\lambda_3)  & = & \bar{v}_{\lambda_1}(p_1) \gamma^\mu 
    v_{\lambda_3}(p_3), \\
    j_2^\mu(\lambda_2,\lambda_4)  & = & \bar{u}_{\lambda_4}(p_4) \gamma^\mu 
    u_{\lambda_2}(p_2),
  \end{array}
  \label{CurLab}
\end{equation}
satisfying current-conservation:
\begin{equation}
  j_i(\lambda_m,\lambda_n) \cdot k_i = 0 \hspace{1.0cm} 
  (i=1,2).
\end{equation}
The polarization vector of the
outgoing spin-1 resonance with
helicity $\lambda_R$ is $\varepsilon^{\mu *}(\lambda_R)$
whereas 
 $\varepsilon^{\mu\nu *}(\lambda_R)$ is the polarization
tensor of the outgoing spin-2 resonances with helicity $\lambda_R$. 
We use the polarization vectors as defined
in (3.15) of \cite{Dittmaier}. With this choice, the
spin-2 polarization tensors can be constructed from the 
spin-1 polarization vectors 
using the Clebsch-Gordan  coefficients
\begin{equation}
  \begin{array}{lcl}
    \varepsilon^{*\mu\nu}(\pm 2) & = & \varepsilon^{*\mu}(\pm 1)
    \varepsilon^{*\nu}(\pm 1), \\
    \varepsilon^{*\mu\nu}(\pm 1) & = & \pm \frac{1}{\sqrt{2}} \left(
    \varepsilon^{*\mu}(\pm 1)\varepsilon^{*\nu}(0) +
    \varepsilon^{*\mu}(0)\varepsilon^{*\nu}(\pm 1) \right), \\
    \varepsilon^{*\mu\nu}(0) & = & \frac{1}{\sqrt{6}} \left(
    -\varepsilon^{*\mu}(+1)\varepsilon^{*\nu}(-1)+2\varepsilon^{*\mu}(0)
    \varepsilon^{*\nu}(0)-\varepsilon^{*\mu}(-1)\varepsilon^{*\nu}(1)
    \right).
  \end{array}
  \label{CGDecomp}
\end{equation}
Finally, the constants $c_1,\ldots,c_5$ are given by 
\begin{equation}
  c_1=g_0, \hspace{0.4cm}
  c_2=4g_1/W, \hspace{0.4cm}
  c_3=2\sqrt{6} g_1, \hspace{0.4cm}
  c_4=4\sqrt{3}Wg_1, \hspace{0.4cm}
  c_5=8\sqrt{30}g_2,
\end{equation}
where we have introduced
\begin{equation}
  g_i=\frac{16 e^2 e_q^2|{\cal R}^{(i)}(0)|\alpha}{(s+s'+u+u'-8m_e^2)^{i+1}}
      \sqrt{\frac{3\pi}{W}}.
\end{equation}
The values used for the fractional charges $e_q^2$ and for (the 
derivatives of) the radial part
of the wave functions in the origin, $|R(0)|$,
can be found in \cite{SchuBerGul}.

The above formulae 
lead to
expressions and numerical results for the total 
matrix element squared and the density matrices. We thus follow a
calculational strategy that differs from the usual one 
\cite{Budnev}, in which  the 
expression for the total 
cross section is split into three parts: two density matrices for the 
production of the virtual photons from the
external leptons and a cross section for the process 
$\gamma^*\gamma^* \to R$, 
which in turn can be written in the well-known form
\renewcommand{\arraystretch}{1.9}
\begin{equation} 
  \begin{array}{lcl}
    \mbox{d}\sigma & = & \frac{\alpha^2}{16\pi^4k_1^2k_2^2}
    \sqrt{\frac{X_{\gamma\gamma}}{X_{e^+e^-}}} 
    \left\{ 4\rho_1^{++}\rho_2^{++}\sigma_{TT} +
    2\rho_1^{++}\rho_2^{00}\sigma_{TS}  \right. \\ &&
    +2\rho_1^{00}\rho_2^{++}\sigma_{ST} +
    \rho_1^{00}\rho_2^{00}\sigma_{SS} + 
    2|\rho_1^{+-}\rho_2^{+-}|\tau_{TT} \cos (2\tilde{\phi})  \\ && \left. 
    -8|\rho_1^{+0}\rho_2^{+0}|\tau_{TS} \cos (\tilde{\phi}) \right\}
    \frac{\mbox{d}^3\vec{{\displaystyle p}}_3}{E_3}
    \frac{\mbox{d}^3\vec{{\displaystyle p}}_4}
   {E_4},
  \end{array}
  \label{BudnevFormula}
\end{equation}
\renewcommand{\arraystretch}{1.1}
where $\tilde{\phi}$ is the angle between the two scattering planes
in the $\gamma\gamma$ rest frame.

In GaGaRes there are three methods implemented for the calculation
of  the square of the 
matrix element summed
over the helicities of the external particles. 

In the first method we
use an expression for the total matrix element
squared in terms of the invariants introduced in (\ref{Invariants}). 
These expressions have been obtained using FORM
\cite{Vermaseren}.  
In the calculation
we have introduced the tensors 
formed by the product of a lepton-current with its complex conjugated, 
summed over all helicities
\begin{equation}
  \begin{array}{lclcl}
    L_1^{\mu\nu} & = & \sum_{\lambda_1,\lambda_3} j_1^\mu(\lambda_1,\lambda_3)
    j_1^{*\nu}(\lambda_1,\lambda_3) & = & 4(p_1^\mu p_3^\nu
    +p_3^\mu p_1^\nu +\frac{1}{2} t g^{\mu\nu} ), \\
    L_2^{\mu\nu} & = & \sum_{\lambda_2,\lambda_4} j_2^\mu(\lambda_2,\lambda_4)
    j_2^{*\nu}(\lambda_2,\lambda_4) & = & 4(p_2^\mu p_4^\nu
    +p_4^\mu p_2^\nu +\frac{1}{2} t' g^{\mu\nu} ). 
  \end{array}
\end{equation}
We also use the standard summation rules for the polarization vectors/tensors
of massive particles
\begin{equation}
  \begin{array}{lcl}
    \sum_{\lambda_R} \varepsilon^\mu (\lambda_R) \varepsilon^{*\nu} (\lambda_R)
    &=& - g^{\mu\nu} + \frac{P_\mu P_\nu}{M^2} \equiv {\cal P}_{\mu\nu},\\
    \sum_{\lambda_R} \varepsilon^{\mu\nu} (\lambda_R) 
    \varepsilon^{*\rho\sigma} (\lambda_R) 
    &=& \frac{1}{2}({\cal P}_{\mu\rho}{\cal P}_{\nu\sigma} +
    {\cal P}_{\mu\sigma}{\cal P}_{\nu\rho} )
    -\frac{1}{3}{\cal P}_{\mu\nu}{\cal P}_{\rho\sigma}.
  \end{array}
\end{equation}
The results of this exercise are collected in 
appendix \ref{MatExs}. The numerical calculation of the total cross
section involving these expressions
is the fastest method in GaGaRes.

We now turn to the two other calculational methods, which are options
in the program. They are based on the evaluation of helicity amplitudes 
in the Weyl-van-der-Waerden (WvdW) spinor formalism 
\cite{Dittmaier,BerendsGiele}.
These methods are slower than the method
using the expression in terms of invariants,
but they allow a straightforward  construction of  density matrices for 
the produced resonances.
In the WvdW calculations we use the notation and conventions introduced in
\cite{Dittmaier}. Furthermore 
the Levi-Civita tensor can be written in the 
WvdW formalism as \cite{Pirani}
\begin{equation}
  \varepsilon_{\dot{A}W\dot{B}X\dot{C}Y\dot{D}Z} = 4{\mbox{i}} \left(
  \varepsilon_{{\dot{A}}{\dot{B}}}\varepsilon_{
  \dot{C}\dot{D}}
  \varepsilon_{
  \vspace{0.5cm}
  WY
  \vspace{-0.5cm}
  } \varepsilon_{XZ} -
  \varepsilon_{\dot{A}\dot{C}}\varepsilon_{\dot{B}\dot{D}}
  \varepsilon_{WX} \varepsilon_{YZ} \right),
\end{equation}
where $\varepsilon_{AB}$ is the metric tensor for the Weyl spinors, given by
\begin{equation}
  \varepsilon_{AB} = \varepsilon_{\dot{A}\dot{B}}=\varepsilon^{AB} = \varepsilon^{\dot{A}\dot{B}}= \left(
    \begin{array}{cc}
      0 & 1 \\
      -1 & 0
    \end{array}
  \right).
\end{equation}
This metric tensor also defines the (antisymmetric) inner product of 
two Weyl spinors
\begin{equation}
  < p k > = \varepsilon^{AB} p_A k_B = p_A k^A = - <k p >, \hspace{0.5cm} < p p > = 0 .
  \label{WvdWsi}
\end{equation}
In the WvdW formalism each Minkowski four-vector $k^\mu$ can be related to a WvdW bispinor
\begin{equation}
  k^\mu=(k^0,k^1,k^2,k^3) \leftrightarrow K_{\dot{A}B} = 
  \sigma_{\mu,\dot{A}B}k^\mu =
  \left(
  \begin{array}{cc}
    k^0 + k^3 & k^1 + \mbox{i}k^2 \\
    k^1 - \mbox{i}k^2 & k^0 - k^3 
  \end{array}
  \right).
\end{equation}
As a consequence the inner 
products of two Minkowski four-vectors can be written as
\begin{equation}
  k \cdot p = \frac{1}{2} K_{\dot{A}B} P^{\dot{A}B} = \frac{1}{2} K^{\dot{A}B} P_{\dot{A}B}, \hspace{0.5cm}
  k^2\delta_B^C = K_{\dot{A}B} K^{\dot{A}C}.
\end{equation}
A WvdW bispinor corresponding to a time-like ($k^2>0$) or light-like 
($k^2=0$) Minkowski four-vector
$k^\mu=(k^0,|\vec{k}|\sin\theta\cos\phi,|\vec{k}|\sin\theta\sin\phi,|\vec{k}|\cos\theta)$ can be 
decomposed into two Weyl spinors 
\begin{equation}
  K_{\dot{A}B} = \sum_{i=1,2} \kappa_{i,\dot{A}} \kappa_{i,B},
\end{equation}
with
\begin{equation}
    \kappa_{1,A}  =  \sqrt{k^0 + |\vec{k}|} \left( 
                       \begin{array}{c}
                         e^{-\mbox{i}\phi} \cos \frac{\theta}{2} \\
                         \sin \frac{\theta}{2}
                       \end{array} \right), 
    \hspace{0.5cm}
    \kappa_{2,A}  =  \sqrt{k^0 - |\vec{k}|} \left( 
                       \begin{array}{c}
                         \sin \frac{\theta}{2} \\
                         -e^{\mbox{i}\phi} \cos \frac{\theta}{2} 
                        \end{array} \right).
\end{equation}
From these formulae one sees that for light-like four-vectors the WvdW expressions simplify dramatically
as for these four-vectors the Weyl spinors with subindex $2$ vanish. For these four-vectors the 
corresponding WvdW bispinor can be written in a simple dyad form.

The usual Dirac spinors for helicity $\pm \frac{1}{2}$ particles and 
antiparticles can also be expressed in the above Weyl spinors
\begin{equation}
  u_+ = \left(
  \begin{array}{c}
    \kappa_{1,A} \\
    -\kappa_2^{\dot{A}}
  \end{array}
  \right), \hspace{0.4cm},
  u_- = \left(
  \begin{array}{c}
    \kappa_{2,A} \\
    \kappa_1^{\dot{A}}
  \end{array}
  \right), \hspace{0.4cm},
  v_+ = \left(
  \begin{array}{c}
    -\kappa_{2,A} \\
    \kappa_1^{\dot{A}}
  \end{array}
  \right), \hspace{0.4cm},
  v_- = \left(
  \begin{array}{c}
    \kappa_{1,A} \\
    \kappa_2^{\dot{A}}
  \end{array}
  \right),
\end{equation}
\begin{equation}
  \bar{u}_+ = ( -\kappa_2^A,\kappa_{1,\dot{A}}), \hspace{0.4cm}
  \bar{u}_- = ( \kappa_1^A,\kappa_{2,\dot{A}}), \hspace{0.4cm}
  \bar{v}_+ = ( \kappa_1^A,-\kappa_{2,\dot{A}}), \hspace{0.4cm}
  \bar{v}_- = ( \kappa_2^A,\kappa_{1,\dot{A}}).
\end{equation}
Using 
\begin{equation}
  \gamma^\mu =
  \left(
  \begin{array}{cc}
    0 & \sigma_{\dot{B}A}^\mu \\
    \sigma_{\dot{A}B}^\mu & 0
  \end{array}
  \right),
\end{equation}
the currents (\ref{CurLab}) can also be written in the WvdW formalism, 
for positrons one has
\renewcommand{\arraystretch}{1.8} 
\begin{equation}
  \begin{array}{lcl}
    J_1^{\dot{R}S}(++) & = & 2\left[ (p_3)_1^{\dot{R}}(p_1)_1^S +
                             (p_1)_2^{\dot{R}}(p_3)_2^S \right],\\
    J_1^{\dot{R}S}(+-) & = & 2\left[ (p_3)_2^{\dot{R}}(p_1)_1^S -
                             (p_1)_2^{\dot{R}}(p_3)_1^S \right],\\
    J_1^{\dot{R}S}(-+) & = & 2\left[ (p_3)_1^{\dot{R}}(p_1)_2^S -
                             (p_1)_1^{\dot{R}}(p_3)_2^S \right],\\
    J_1^{\dot{R}S}(--) & = & 2\left[ (p_1)_1^{\dot{R}}(p_3)_1^S +
                             (p_3)_2^{\dot{R}}(p_1)_2^S \right].\\
  \end{array}
\end{equation}
\renewcommand{\arraystretch}{1.1}
The currents for the electron line can be obtained from these currents by
making the substitutions $p_1 \leftrightarrow p_2$ and 
$p_3 \leftrightarrow p_4$. 

It can be checked that these WvdW bispinors 
satisfy current-conservation, which reads in WvdW notation
\begin{equation}
  J_i^{\dot{A}B}(\lambda_m,\lambda_n) K_{i,\dot{A}B} = 0 \hspace{1.0cm}
  (i=1,2).
\end{equation}

Finally, the polarization tensors for the
massive spin-2 resonances can be constructed from the spin-1 polarization 
vectors using the Clebsch-Gordan coefficients given in (\ref{CGDecomp}).

The expressions for the amplitudes in 
(\ref{StandardAmplitudes}) read in the WvdW notation
\begin{equation}
  \begin{array}{lcl}
     {\cal M}(^1S_0)
     (\lambda_1,\lambda_2,\lambda_3,\lambda_4)
     & = &
     \frac{c_1}{4tt'} \left\{ J_{1,\dot{C}Y}(\lambda_1,
     \lambda_3)K_1^{\dot{B}Y}K_{2,\dot{B}Z}
     J_2^{\dot{C}Z}(\lambda_2,\lambda_4) \right. \\
     && \left. -J_{1,\dot{C}Y}(\lambda_1,
     \lambda_3)K_1^{\dot{C}X}K_{2,\dot{D}X}
     J_2^{\dot{D}Y}(\lambda_2,\lambda_4) \right\}, \\
     {\cal M}(^3P_0)
     (\lambda_1,\lambda_2,\lambda_3,\lambda_4)
     & = & \frac{c_2}{4tt'} \left\{
     2H J_{1,\dot{A}B}(\lambda_1,\lambda_3)
      J_2^{\dot{A}B}(\lambda_2,\lambda_4) \right. \\
     && \left.
     -G J_{1,\dot{A}B}(\lambda_1,\lambda_3)
     J_{2,\dot{C}D}(\lambda_2,\lambda_4)
     K_2^{\dot{A}B}K_1^{\dot{C}D} \right\}, \\
     {\cal M}(^3P_1)
     (\lambda_1,\lambda_2,\lambda_3,\lambda_4,\lambda_R)
     & = & \frac{c_3}{4tt'} \left\{t[J_{2,\dot{C}Y}(\lambda_2,\lambda_4)
     \varepsilon^{*\dot{B}Y}(\lambda_R) J_{1,\dot{B}Z}(\lambda_1,\lambda_3)
     K_2^{\dot{C}Z} \right. \\ &&
      -J_{2,\dot{C}Y}(\lambda_2,\lambda_4)
     \varepsilon^{*\dot{C}X}(\lambda_R) J_{1,\dot{D}X}(\lambda_1,\lambda_3)
     K_2^{\dot{D}Y}] \\
     && +t' [ J_{1,\dot{C}Y}(\lambda_1,\lambda_3)
     \varepsilon^{*\dot{B}Y}(\lambda_R) J_{2,\dot{B}Z}(\lambda_2,\lambda_4)
     K_1^{\dot{C}Z} \\ &&
     \left. -
     J_{1,\dot{C}Y}(\lambda_1,\lambda_3)
     \varepsilon^{*\dot{C}X}(\lambda_R) J_{2,\dot{D}X}(\lambda_2,\lambda_4)
     K_1^{\dot{D}Y}]
    \right\}, \\
    {\cal M}(^3P_2)
     (\lambda_1,\lambda_2,\lambda_3,\lambda_4,\lambda_R)
     & = &  \frac{c_4}{8tt'} \left\{
     2FJ_{1,\dot{A}B}(\lambda_1,\lambda_3)J_{2,\dot{C}D}(\lambda_2,\lambda_4)
     \right. \\ &&
     +K_{1,\dot{A}B}K_{2,\dot{C}D}
     J_{1,\dot{E}F}(\lambda_1,\lambda_3)J_2^{\dot{E}F}(\lambda_2,\lambda_4) 
     \\
     && -K_{1,\dot{A}B}J_{2,\dot{C}D}(\lambda_2,\lambda_4) 
     J_{1,\dot{E}F}(\lambda_1,\lambda_3)K_2^{\dot{E}F}
     \\ && \left.
     -K_{2,\dot{A}B}J_{1,\dot{C}D}(\lambda_1,\lambda_3) 
     J_{2,\dot{E}F}(\lambda_2,\lambda_4)K_1^{\dot{E}F}
    \right\}
     \\ && \varepsilon^{*\dot{A}B\dot{C}D}(\lambda_R), \\
     {\cal M}(^1D_2)
     (\lambda_1,\lambda_2,\lambda_3,\lambda_4,\lambda_R)
     & = & \frac{c_5}{4c_1} \left\{K_{1,\dot{A}B} K_{2,\dot{C}D}
     \varepsilon^{*\dot{A}B\dot{C}D}(\lambda_R)\right\} \\ &&
     {\cal M}(^1S_0)
     (\lambda_1,\lambda_2,\lambda_3,\lambda_4),
  \end{array}
  \label{WvdWAmplitudes}
\end{equation}
where
\begin{equation}
  \begin{array}{lcl}
    F & = & k_1 \cdot k_2 , \\
    G & = & W^2+F , \\
    H & = & FG-tt'.
  \end{array}
\end{equation}

The second and third calculational methods in GaGaRes both 
use the WvdW calculation.
In the second  method 
the matrix elements are finally expressed  in terms of the standard 
spinor inner products as defined in equation (\ref{WvdWsi}). 
For the derivation of these expressions FORM has been used.

The third method expresses the matrix element in traces of $2 \times 2$ 
matrices.
Let us introduce 
a down-matrix $K_\downarrow$
\begin{equation}
  K_\downarrow \equiv K_{\dot{A}B},
\end{equation}
and an up-matrix $K^\uparrow$
\begin{equation}
  K^\uparrow \equiv (K_\uparrow)^T, \hspace{2.0cm} K_\uparrow \equiv 
  K^{\dot{A}B}.
\end{equation}
With these matrices a product of an even number of WvdW bispinors can be 
rewritten as the trace of a product of the previously defined 
down- and up-matrices, e.g.
\begin{equation}
  \begin{array}{lcl}
    P^{\dot{A}B}K_{\dot{A}B} & = & \mbox{Tr}\left[ P^\uparrow K_\downarrow 
    \right] = 2 p \cdot k, \\
    P_{\dot{A}B}Q^{\dot{C}B} R_{\dot{C}D}S^{\dot{A}D} & = &
    \mbox{Tr}\left[P_\downarrow Q^\uparrow R_\downarrow S^\uparrow \right]. 
  \end{array}
\end{equation}
One also needs the expressions for the spin-2 polarization vectors in this
method
\begin{equation}
  \begin{array}{lcl}
    \mbox{Tr}\left[  P^\uparrow\varepsilon_{
    \downarrow\downarrow}^*(\pm 2)  K^\uparrow 
    \right] & = & 
    \mbox{Tr}\left[ P^\uparrow 
    \varepsilon_\downarrow^*(\pm 1)] \right]
    \mbox{Tr}\left[ K^\uparrow 
    \varepsilon_\downarrow^*(\pm 1)] \right],
    \\
    \mbox{Tr}\left[  P^\uparrow\varepsilon_{
    \downarrow\downarrow}^*(\pm 1)  K^\uparrow 
    \right] & = & \frac{ \pm 1}{\sqrt{2}} \left( 
    \mbox{Tr}\left[ P^\uparrow 
    \varepsilon_\downarrow^*(\pm 1) \right]
    \mbox{Tr}\left[ K^\uparrow 
    \varepsilon_\downarrow^*(0) \right] \right. \\
    && \left. +
    \mbox{Tr}\left[ P^\uparrow 
    \varepsilon_\downarrow^*(0) \right]
    \mbox{Tr}\left[ K^\uparrow 
    \varepsilon_\downarrow^*(\pm 1) \right]
    \right),
    \\
    \mbox{Tr}\left[  P^\uparrow\varepsilon_{
    \downarrow\downarrow}^*(0)  K^\uparrow 
    \right] & = & \frac{ 1}{\sqrt{6}} \left( 
    -\mbox{Tr}\left[ P^\uparrow 
    \varepsilon_\downarrow^*(+ 1) \right]
    \mbox{Tr}\left[ K^\uparrow 
    \varepsilon_\downarrow^*(- 1) \right]  \right. \\
     &  & 
    +
    2 \mbox{Tr}\left[ P^\uparrow 
    \varepsilon_\downarrow^*(0) \right]
    \mbox{Tr}\left[ K^\uparrow 
    \varepsilon_\downarrow^*(0) \right] \\
    && \left. -
    \mbox{Tr}\left[ P^\uparrow 
    \varepsilon_\downarrow^*(- 1) \right]
    \mbox{Tr}\left[ K^\uparrow 
    \varepsilon_\downarrow^*(+ 1) \right]
    \right) .
  \end{array}
\end{equation}
As a result the matrix elements can be written as 
\begin{equation}
  \begin{array}{lcl}
   {\cal M}(^1S_0) & = & \frac{c_1}{4tt'} \mbox{Tr} \left[
    J_{1\downarrow}(\lambda_1,\lambda_3) \left\{
    K_1^\uparrow K_{2\downarrow} J_2^\uparrow(\lambda_2,
    \lambda_4) \right. \right. \\
    && \left. \left.
     -  J_2^\uparrow(\lambda_2,\lambda_4)  K_{2\downarrow}
     K_1^\uparrow \right\} \right], \\
    {\cal M}(^3P_0) & = & \frac{c_2}{4tt'}
    \left( 2H\mbox{Tr}\left[ J_{1\downarrow}(\lambda_1,
    \lambda_3) J_2^\uparrow(\lambda_2,\lambda_4) \right] \right. \\
    && \left.
    -G \mbox{Tr} \left[  J_{1\downarrow}(\lambda_1,
    \lambda_3) K_2^\uparrow \right]
     \mbox{Tr}\left[  J_{2\downarrow}(\lambda_2,
    \lambda_4) K_1^\uparrow \right] \right), \\
    {\cal M}(^3P_1) & = & \frac{c_3}{4tt'} \left(
    t \mbox{Tr}\left[ J_2^\uparrow(\lambda_2,
    \lambda_4) \left\{ \varepsilon_{\downarrow}^*
    (\lambda_R) J_1^\uparrow(\lambda_1,
    \lambda_3)   K_{2\downarrow} \right. \right. \right. \\
    && \left. \left.-  K_{2\downarrow}
    J_1^\uparrow(\lambda_1,
    \lambda_3) \varepsilon_{\downarrow}^*
    (\lambda_R) \right\} \right] \\
    && 
    + t'\mbox{Tr}\left[   J_1^\uparrow(\lambda_1,
    \lambda_3) \left\{ \varepsilon_{\downarrow}^*
    (\lambda_R) J_2^\uparrow(\lambda_2,
    \lambda_4) K_{1\downarrow}
    \right. \right. \\ 
    && \left. \left. \left.
    - K_{1\downarrow}
    J_2^\uparrow(\lambda_2,
    \lambda_4)\varepsilon_{\downarrow}^*
    (\lambda_R) \right\} \right] \right), \\
    {\cal M}(^3P_2) & = & \frac{c_4}{8tt'} \left(
    2F \mbox{Tr}\left[  J_1^\uparrow(\lambda_1,
    \lambda_3) \varepsilon_{\downarrow\downarrow}^*
    (\lambda_R)  J_2^\uparrow(\lambda_2,
    \lambda_4) \right] \right. \\ 
    && +
    \mbox{Tr}\left[  K_1^\uparrow\varepsilon_{
    \downarrow\downarrow}^*(\lambda_R)  K_2^\uparrow 
    \right] 
    \mbox{Tr}\left[ J_{1\downarrow}(\lambda_1,
    \lambda_3) J_2^\uparrow(\lambda_2,
    \lambda_4) \right]   
    \\
    && 
    -\mbox{Tr}\left[  K_1^\uparrow\varepsilon_{
    \downarrow\downarrow}^*(\lambda_R) 
    J_2^\uparrow(\lambda_2,\lambda_4)  \right]
    \mbox{Tr}\left[  J_{1\downarrow}(\lambda_1,
    \lambda_3)    K_2^\uparrow  \right]  
    \\ && \left.
    -\mbox{Tr}\left[  K_2^\uparrow\varepsilon_{
    \downarrow\downarrow}^*(\lambda_R) 
    J_1^\uparrow(\lambda_1,\lambda_3)  \right]
    \mbox{Tr}\left[  J_{2\downarrow}(\lambda_2,
    \lambda_4)    K_1^\uparrow  \right]  \right), \\
    {\cal M}(^1D_2) & = & \frac{c_5}{4c_1}
    \mbox{Tr}\left[  K_1^\uparrow\varepsilon_{
    \downarrow\downarrow}^*(\lambda_R)  K_2^\uparrow 
    \right] 
    {\cal M}(^1S_0).    
  \end{array}
  \label{MatMat}
\end{equation}
It turns out that this is the fastest method using the 
WvdW formalism, thanks to a fast matrix multiplication in FORTRAN90. 
The other method using the WvdW 
spinorial inner products
merely serves as a cross-check.

For the evaluation of the matrix element squared or of the density matrices, 
summations over the helicities of external particles have to be made. However,
the matrix element need not be evaluated for every helicity configuration.
From the choice of the polarization vectors in \cite{Dittmaier} it can be
shown that the complex conjugate of a matrix element is related to 
the matrix element with flipped helicities:
\begin{equation}
  \begin{array}{llcl}
    \mbox{Spin-0 resonances :} & {\cal M}(\{-\lambda_i\}) & = & \left( \prod_i
    \lambda_i \right) {\cal M}^*(\{ \lambda_i\}), \\
    \mbox{Spin-1 resonance :} & {\cal M}(\{-\lambda_i\},-\lambda_R) & = & 
    \left( \prod_i
    \lambda_i \right)
    {\cal M}^*(\{ \lambda_i\},\lambda_R), \\
    \mbox{Spin-2 resonances :} & {\cal M}(\{-\lambda_i\},-\lambda_R) & = & 
    \left( \prod_i
    \lambda_i \right) 
    (-1)^{\lambda_R} 
    {\cal M}^*(\{ \lambda_i\},\lambda_R). 
  \end{array}
  \label{MatSym}
\end{equation}
Here again the $\lambda_i$'s are referring to the incoming and outgoing 
leptons and $\lambda_R$ is the helicity of the resonance.


It should be noted that in the limit $m_e\to 0$, where the WvdW calculations
simplify  dramatically,
we have analytically checked
that the WvdW calculation gives the same result for the total matrix element 
squared as the massless parts of the cross sections in appendix 
\ref{MatExs}.


\section{Density Matrices}
When one considers the
decay of a resonance into some final state $X$,
\begin{equation}
  e^+e^- \to e^+e^-R \to e^+ e^- X,
  \label{ResDecay}
\end{equation}
one can write for the
amplitude 
\begin{equation}
  {\cal M} = \sum_{\lambda_R} \xi(P,W) 
  {\cal A}_{\lambda_R} {\cal D}_{\lambda_R}.
\end{equation}
${\cal A}_{\lambda_R}$ describes the (two-photon) production of a resonance
with helicity $\lambda_R$. This is our previously discussed matrix element.
${\cal D}_{\lambda_R}$ describes the decay of 
a resonance with helicity $\lambda_R$ into the final state $X$.
The quantity $\xi(P,W)$ represents the propagator of the resonance
and numerical factors. 
The total matrix element still depends on the external four-momenta. 
No implicit integrations over outgoing momenta have been carried out.

The square of the total matrix element is
then given by
\begin{equation}
  \sum |{\cal M}|^2  
  = |\xi(P,M)|^2 \sum_{\lambda_R,\lambda_{R}'} {\cal A}_{\lambda_R \lambda_R'} 
  {\cal D}_{\lambda_R \lambda_R'}
  = |\xi(P,M)|^2 \mbox{Tr}\left({\cal A} {\cal D}^* \right).
\end{equation}
Here $\sum$ is the summation over the initial and final states and the
helicities of the intermediate state. The quantities 
${\cal A}_{\lambda_R \lambda_R'}$ and
${\cal D}_{\lambda_R \lambda_R'}$ are
${\cal A}_{\lambda_R}{\cal A}_{\lambda_R'}^*$
and ${\cal D}_{\lambda_R} {\cal D}_{\lambda_R'}^*$, summed over 
the helicities of all particles but the resonance.
%
They are the
density matrices for the production and decay  of the 
resonance. For a spin $J$ resonance the density matrix is formed by
a $(2J+1) \times (2J+1)$-matrix.

GaGaRes calculates for every generated event the 
normalized density matrix for the production of the resonance
\begin{equation}
  \rho_{\lambda_R \lambda_R'} = \frac{\sum_{\lambda_i} {\cal A}( \lambda_i,
  \lambda_R) {\cal A}^*( \lambda_i,\lambda_R')}
  {\sum_{\lambda_i,\lambda_R} |{\cal A}( \lambda_i,
  \lambda_R)|^2},   
  \label{NorMat}
\end{equation}
where the summation in the numerator is over the electron and positron
helicities and in the denominator the summation is over all 
helicities.

The density matrix for the production only depends on the production
process and can therefore be calculated in GaGaRes. If one knows the 
mechanism for the decay of the resonance into a state $X$, also that
decay density matrix could be incorporated in GaGaRes. This is beyond the
scope of the present investigation.

From the definition of $\rho$ one can 
immediately derive two properties of this 
density matrix:
\begin{itemize}
  \item $\rho$ is hermitian : $\rho=\rho^\dagger$;
  \item The trace of the density matrix equals $1$ : $\mbox{Tr}\left(
  \rho \right) =1$.
\end{itemize}

However, from the relation between complex conjugated matrix elements
and spin flipped matrix elements  (\ref{MatSym}) and the definition 
for the normalized 
density matrix (\ref{NorMat}), one can also 
derive an additional symmetry. For spin-1 resonances it reads
\begin{equation}
  \rho_{\lambda_R\lambda_R'} =  \rho_{-\lambda_R' -\lambda_R}.
\label{ExtraSym1}  
\end{equation}
\begin{equation}
  \rho_{\lambda_R\lambda_R'} =  (-1)^{\lambda_R+\lambda_R'} \rho_{-\lambda_R' -\lambda_R}.
\label{ExtraSym2}  
\end{equation}

This symmetry, the hermiticity and the fact that the trace is one, 
reduce strongly the number of 
independent elements in the density matrix. For a spin-1 resonance
the density matrix has one real diagonal element and
two complex off-diagonal elements, e.g.~$\rho_{-1,1}$ and $\rho_{-1,0}$.
For a spin-2 resonance one has two real diagonal elements 
and 6 independent complex off-diagonal elements. The fact that 
the diagonal elements are real follows from hermiticity.


In GaGaRes only the relations in (\ref{MatSym}) have  been used to 
reduce the amount of calculations. However, checks have been performed
that the density matrices are hermitian and that equations
(\ref{ExtraSym1}) and (\ref{ExtraSym2})
hold.

The density matrices
are computed for every event
in the resonance rest frame (RRF), as this is the most natural 
scheme for the description of the resonance decay. 
The positive $z$-axis, the quantization axis, is defined by the 
boost direction. The 
Lorentz-transformation from the lab-frame to this RRF is performed by
subsequently
an azimuthal rotation, a polar rotation and a boost.
The rotation brings the momentum of the resonance in the lab frame
along the $z$-axis and the boost then brings the resonance to rest.
From an experimental point of view this choice of polarization axis is also
favoured as it is straightforward to reconstruct the boost direction from
the detected final state $X$.

Some results on the calculation of density matrices using GaGaRes and a 
discussion thereof can be found in \cite{Gulik}.


\section{MC Generation Schemes}
The parametrization of the external four-momenta in the lab-frame is given
in (\ref{MomPar}). From now
on we will use the following shorthand notations
\begin{equation} 
  \begin{array}{lcl}
    s_i & \equiv & \sin \theta_i, \\
    c_i & \equiv & \cos \theta_i.
  \end{array}
\end{equation}
Additionally we introduce $\theta_{34}$, the angle between the two outgoing
leptons, given by
\begin{equation}
  c_{34} \equiv \cos \theta_{34} = - c_3 c_4 - s_3 s_4 \cos(\phi_3-\phi_4+\pi).
\end{equation}

The general expression for the cross section of a $2 \to 3$ process 
reads
\begin{equation}
  \begin{array}{lcl}
    \sigma & = &  {\displaystyle \int} 
    (2\pi)^4 \delta^{(4)}(p_1+p_2-p_3-p_4-P) 
    \frac{1}{4\sqrt{(p_1 \cdot p_2)^2-m_e^4}} \frac{1}{4} \sum |{\cal M}|^2 \\
     &&\frac{\mbox{d}^3\vec{{\displaystyle p}}_3}{(2\pi)^32E_3}
     \frac{\mbox{d}^3\vec{{\displaystyle p}}_4}{(2\pi)^32E_4}
     \frac{\mbox{d}^3\vec{{\displaystyle P}}}{(2\pi)^32E_R}.
  \end{array}
\end{equation}
Performing the integration over $\vec{P}$ and rewriting the remaining delta
function gives
\begin{equation}
  \begin{array}{lcl}
    \sigma & = & {\displaystyle \int} 
    \frac{1}{(2\pi)^5} \frac{\frac{1}{4}\sum |{\cal M}|^2}
    {8sE_3E_4\sqrt{1-\frac{4m_e^2}{s}}}
    \delta(4E_b^2-4E_b(E_3+E_4)+2m_e^2+2E_3E_4 \\
    && -2|\vec{p}_3| | \vec{p}_4| \cos \theta_{34}
     - M^2)  \mbox{d}^3 \vec{p}_3 \mbox{d}^3 \vec{p}_4.
  \end{array}
\end{equation}
Solving the 
delta function for $E_3$ yields
\begin{equation}
  E_3 = \frac{-\xi(E_4-2E_b)\pm {D}}
  {2 \{ (E_4-2E_b)^2-|\vec{p}_4|^2c_{34}^2 \} },
  \label{E3ABC}
\end{equation}
where
\begin{equation}
  \begin{array}{lcl}
     \xi & = & 4E_b^2-4E_bE_4+2m_e^2-M^2, \\
     D   & = & |\vec{p}_4|c_{34} |{\tilde{\Delta}}|, \\
      {\tilde{\Delta}}^2 & = & \xi^2 - 4(E_4-2E_b)^2m_e^2 
      +4|\vec{p}_4|^2c_{34}^2 m_e^2.
  \end{array}
\end{equation}
The sign ambiguity in (\ref{E3ABC}) 
can be solved by dividing the phase-space into 
two parts. The cross sections of the two solutions
are denoted by $\sigma_\pm$, corresponding
to the sign in (\ref{E3ABC}).

For the two solutions we find
\begin{itemize}
  \item $-$ solution
    \begin{equation} 
      \begin{array}{l}
        E_{4,min}=m_e, \\
        E_{4,max}=E_b-\frac{m_eM}{2E_b} -\frac{M^2}{4E_b}, \\
        m_e \leq E_4 \leq E_4' \mbox{ then } 
        -1\leq c_{34} \leq +1, \\
        E_4'\leq E_4 \leq E_{4,max}\mbox{ then } -1
          \leq c_{34} \leq-c_a
      \end{array}
    \end{equation}
  \item $+$ solution
    \begin{equation}
      \begin{array}{l}
        E_{4,min}=E_4', \\
        E_{4,max}=E_b-\frac{m_eM}{2E_b} -\frac{M^2}{4E_b}, \\
        -1 \leq c_{34} \leq -c_a,
      \end{array}
    \end{equation}
\end{itemize}
where 
\begin{equation}
  E_4'=\frac{4E_b^2-4m_eE_b+2m_e^2-M^2}{4E_b-2m_e},
\end{equation}
and
\begin{equation}
  c_a=\sqrt{\frac{4(2E_b-E_4)^2m_e^2-(4E_b^2-4E_4E_b+2m_e^2-M^2)^2}
  {4|\vec{p_4}|^2m_e^2}}.
\end{equation}
Making a change of variables and performing the integration over $E_3$
leads to the total cross section
\begin{equation}
  \sigma_\pm = \int \frac{\frac{1}{4} \sum |{\cal M}|^2}{256\pi^5s}
  \frac{1}{\sqrt{1-\frac{4m_e^2}{s}}}
  \frac{|\vec{p}_3||\vec{p}_4|}{|4E_b-2E_4+2
  \frac{|\vec{p}_4|E_3}{|\vec{p}_3|}c_{34}|}
  \mbox{d} E_4 \mbox{d}^2 \Omega_3 \mbox{d}^2 \Omega_4.
\end{equation}

From now on all all variables are normalized to the beam energy $E_b$.
After this normalization the cross section reads
\begin{equation}
  \sigma_\pm = \int \frac{\frac{1}{4} \sum|{\cal M}|^2}{256\pi^5s}
  \frac{1}{\sqrt{1-\frac{4m_e^2}{s}}}
  \frac{|\vec{p}_3||\vec{p}_4|}{|4-2E_4+2
  \frac{|\vec{p}_4|E_3}{|\vec{p}_3|}c_{34}|}
  \mbox{d} E_4 \mbox{d}^2 \Omega_3 \mbox{d}^2 \Omega_4.
  \label{NormSig}
\end{equation}

GaGaRes offers the user the choice between two schemes
in which different sets of variables are generated to perform the
MC integration.
Scheme I, based on  
\cite{DavBerKlei},  is the most efficient scheme for 
no-tag events, as the peaking behaviour of the matrix element is better 
described. 
Scheme II is more efficient for single- 
or double-tag events, as both polar angles are generated variables.

Scheme I is described in \cite{DavBerKlei}
for the process
\[
e^+e^- \to e^+e^-l^+l^- \hspace{0.3cm}(l=e,\mu,\tau), 
\]
where the outgoing electron and positron are not tagged.
As we have here one particle less in the final state,
some simplifications
can be introduced. 
This generation scheme being very similar to the scheme in
\cite{DavBerKlei}, will not be discussed in this paper.

The second scheme 
is more efficient not only in 
the generation of single- and double-tagged events, but also
for the generation of events with cuts on $Q_i^2$,
where an automatic cut on the angles is imposed.

The procedure is to 
start with the normalized expression given in (\ref{NormSig}). The first
approximation one makes is to replace the expression for the matrix element
\begin{equation} 
  \frac{1}{4} \sum |{\cal M}|^2 \to \frac{64}{tt'}.
\end{equation}   
The corresponding weight factor is given by
\begin{equation}
  W_1 = \frac{tt'\frac{1}{4} \sum |{\cal M}|^2}{64}.
\end{equation}
In this approximation it is better to put a factor $tt'$ rather than a
factor $t^2t'^2$ in the denominator: from the formulae in appendix 
\ref{MatExs} it follows that in the limit where $t$ and $t'$ are
of order $m_e^2$, the numerator also is of order $m_e^2$, for it can be
shown that in this limit the invariant $\sigma_{2-}$ is also proportional to
$m_e^2$. This reduces effectively 
the behaviour of the denominator to $tt'$. This 
approximation also gives the well-known logarithmic behaviour of 
the total cross section with respect to the total energy.

The cross section now reads 
\begin{equation}
  \sigma_\pm = \frac{1}{4\pi^5s} \frac{1}{\sqrt{1-\frac{4m_e^2}{s}}}
  \int \frac{1}{tt'} \frac{|\vec{p}_3||\vec{p}_4|}
  {|4-2E_4+2\frac{|\vec{p}_4|E_3}{|\vec{p}_3|}c_{34}|} d E_4 \mbox{d}^2\Omega_3
  \mbox{d}^2 \Omega_4.
\end{equation}

Before performing the next approximations, let us consider the
photon propagators more closely. The square of the invariant mass of the 
photon, radiated by the
incoming positron, can be written as 
\begin{equation}
  t=-2E_3(1-c_3+\delta_3),
\end{equation}
with
\begin{equation}
  \delta_3 = \left( \frac{1-|\vec{p}_3|P_b}{E_3} \right)c_3-\frac{m_e^2}{E_3}.
\end{equation}
Now $\delta_{3,min}$ is defined as
\begin{equation}
  \delta_{3,min} \left. \equiv \delta_3 \right|_{c_3=1}
  = \frac{m_e^2(1-E_3)^2}{2E_3^2},
\end{equation}
which reaches its minimum at $E_3=E_{3,max}=1-\frac{1}{2}m_eM
-\frac{1}{4}M^2$ and is denoted by
\begin{equation}
  \varepsilon \equiv \left. \delta_{3,min}\right|_{E_3=E_{3,max}} = 
  \frac{m_e^2(m_eM+\frac{1}{2}M^2)^2}{8(1-\frac{1}{2}m_eM-\frac{1}{4}M^2)^2}.
\end{equation}
For $t'$ we find a similar expression with the subscript $3$ replaced by $4$.

The second approximation now consists of the substitution
\begin{equation}
  \frac{1}{tt'} \to \frac{1}{4E_3E_4(1-c_3+\varepsilon)(1-c_4+\varepsilon)}.
\end{equation}
The corresponding weight factor reads
\begin{equation}
  W_2=\frac{4E_3E_4(1-c_3+\varepsilon)(1-c_4+\varepsilon)}{tt'}.
\end{equation}

The third approximation consists of the approximation in the
denominator
\begin{equation}
  \frac{|\vec{p}_3||\vec{p}_4|}{E_3 E_4}
  |2-E_4+\frac{|\vec{p}_4|E_3}{|\vec{p}_3|}c_{34}| \to 1-E_4.
\end{equation}
This approximation works rather well, as for most generated events the
outgoing leptons go (almost) back to back ($c_{34} \approx -1$).
\begin{equation}
  W_3 = \frac{p_3p_4}{E_3E_4} \frac{1-E_4}
  {|2-E_4+\frac{|\vec{p}_4|E_3}{|\vec{p}_3|}c_{34}|}.
\end{equation}

After these approximations the cross section reads
\begin{equation}
  \sigma_\pm = \frac{1}{32\pi^5s} \frac{1}{\sqrt{1-\frac{4m_e^2}{s}}}
  \int \frac{\mbox{d}E_4\mbox{d}^2\Omega_3\mbox{d}^2\Omega_4}
  {(1-c_3+\varepsilon)
  (1-c_4+\varepsilon)(1-E_4)}.
\end{equation}

First, the trivial azimuthal integrations of $\phi_3$ and $\phi_4$ over
$[\phi_{min},\phi_{max}]$ 
can be performed. The integrations over the polar angles and
over $E_4$ can be performed by using the standard integral
\begin{equation}
  \int_{x_{min}}^{x_{max}} \frac{\mbox{d}x}{a-x} = \ln \left|
  \frac{a-x_{min}}{a-x_{max}} \right|.
  \label{OneOverAMinX}
\end{equation}

The results of the integrations are put into the last two weights,
\begin{equation}
  \begin{array}{lcl}
     W_4 & = & \ln \left| \frac{1+\varepsilon-c_{3,min}}
     {1+\varepsilon-c_{3,max}} \right|
     \ln \left| \frac{1+\varepsilon-c_{4,min}}
     {1+\varepsilon-c_{4,max}} \right| (\phi_{3,max}-\phi_{3,min})
     (\phi_{4,max}-\phi_{4,min})
     , \\ && \\
     W_{5,\pm} & = & \ln \left| \frac{1-E_{4,min}}{1-E_{4,max}} \right|,
  \end{array}
\end{equation}
while the ``total cross section'' is given by
\begin{equation}
  \sigma_{tot} = \frac{1}{32\pi^5s} \frac{1}{\sqrt{1-\frac{4m_e^2}{s}}}.
\end{equation}

The real total cross section can now be determined using
\begin{equation}
  \sigma = \sigma_{tot} \left( 
  <W_1W_2W_3W_4(W_{5+}+W_{5-})>
  \right),
\end{equation}
where $<W_1W_2W_3W_4(W_{5+}+W_{5-})>$ is defined as the average of the total
weight. During the generation of events there is a probability
$W_{5-}/(W_{5+}+W_{5-})$ that an event is generated in the phase space
of the $-$ solution. Especially for the heavier resonances almost all
events will be generated in the $-$ part of the 
solution. 

It is useful to give some comments on the generation of the integration 
variables. The variables are now generated by importance sampling in the
following order: $\phi_3$, $\phi_4$, $\cos \theta_3$, $\cos\theta_4$.
Let ${\cal R}$ be a random variable uniformly distributed over
the interval $[0,1]$. Now the azimuthal angles $\phi_3$ and $\phi_4$
are generated according to
\begin{equation}
  \begin{array}{lcl}
    \phi_3 & = & (\phi_{3,max}-\phi_{3,min}){\cal R}+\phi_{3,min}, \\
    \phi_4 & = & (\phi_{4,max}-\phi_{4,min}){\cal R}+\phi_{4,min}.  
  \end{array}
\end{equation}
Variables that are distributed like the integrand of (\ref{OneOverAMinX}) 
should be generated according to
\begin{equation}
  x=a-(a-x_{max})\left(\frac{a-x_{min}}{a-x_{max}} \right)^{\cal R},
\end{equation}
where it should be noted that in the program $1-x$ rather than $x$ is
generated in order to guarantee the best numerical stability.
At this point $\cos\theta_{34}$ can be evaluated. Then $E_4$ is generated.
When $E_4<E_4'$ one has only a contribution from the $-$ solution of the
delta function. When  $E_4>E_4'$ one has to check whether $c_{34}$ is
in the allowed interval. If so, one has contributions from both
solutions of the delta function.

Cuts on the generated variables are incorporated in the generation 
mechanism. They do not lead to a drop in efficiency (exactly the reason
why scheme II is favoured for single- and double-tag reactions). For 
scheme II the program also offers the user the possibility to impose
cuts on $E_3$, on the photon virtualities $Q_i^2$ and 
on $c_{34}$. The cuts are
applied by the following replacement
\begin{equation}
  |{\cal M}|^2 \to |{\cal M}|^2 \prod \theta,
\end{equation}
where $\prod \theta$ is a product of step functions representing the
cuts.

These cuts lead to a drop in generation efficiency. However, for the
cuts on the photon virtualities $Q_i^2$ the angles of the corresponding polar
angles are adapted in order to decrease this drop in efficiency.
For a cut on the virtualities we use the following approximation formulae
\begin{equation}
  \theta_{i,{min}} = 2 \mbox{arcsin} \sqrt{ \frac{Q_{i,{min}}^2
  +2m_e^2}{4E_bE_{3,{max}}}},
\end{equation}
\begin{equation}
  E_{3,{min}} = \frac{Q_{i,{min}}^2+2m_e^2}{4E_b
  \sin^2\left( \frac{\theta_{3,{max}}}{2} \right)},
\end{equation}
where $\theta_i$ is the minimum polar angle of the outgoing lepton corresponding to the 
photon virtuality $Q_i$. On this photon virtuality the cut $Q_i > Q_{i,{min}}$
is imposed.

The average normalized density matrix is calculated by taking the weighted sum
of the normalized density matrix and by dividing this sum by the 
total weight. This means that for every generated event in the lab system
the momenta in the RRF are obtained which are then used to evaluate the 
RRF density matrix.

The program generates at the same time
unweighted events, i.e.\ events with a weight
factor equal to 1. This is done by applying a hit or miss algorithm.


\section{Structure of the Program}
A complete
overview of all subroutines can be found in the code-listing of GaGaRes.

In the main program the file \verb|gamgaminput| is read out. This file
contains some input parameters. The variables that are read from 
\verb|gamgaminput| are marked with an $^*$ in the following discussion. 


\subsection{Modules}
A list of the most important modules in GaGaRes

\begin{verbatim}
  MODULE global_data
\end{verbatim}

In this module general variables are stored. Also some cuts and switches
are contained in this module.

\vspace{0.5cm}
 
\begin{tabular}{ll}
  \verb|Events|$^*$ & Number of events to be generated. \\
  \verb|StartNumber|$^*$  & Number of events to be generated to determine the 
                        \\
                        & maximum weight. \\
  \verb|IAcc|     & Number of accepted events (after hit or miss 
                    integration). \\
  \verb|ICut|     & Number of events after applied cuts. \\
  \verb|Eb|$^*$   & Beam energy. \\
  \verb|EWE|$^*$  & The expected maximum weight. \\
  \verb|Th3Min|$^*$,\verb|Th3Max|$^*$ & The minimum and maximum allowed 
                    polar angle of \\
                  & the scattered positron. 
\end{tabular}

\begin{tabular}{ll}
  \verb|Th4Min|$^*$,\verb|Th4Max|$^*$ & The minimum and maximum allowed 
                    polar angle of \\
                  & the scattered electron. \\ 
  \verb|Ph3Min|$^*$,\verb|Ph3Max|$^*$ & The minimum and maximum allowed 
                    azimuthal angle of \\
                  & the scattered positron. \\
  \verb|Ph4Min|$^*$,\verb|Ph4Max|$^*$ & The minimum and maximum allowed 
                    azimuthal angle of \\
                  & the scattered electron. \\ 
  \verb|E3Min|$^*$,\verb|E3Max|$^*$ & The minimum and maximum allowed 
                    energy of the \\
                  & scattered positron. \\
  \verb|E4Min|$^*$,\verb|E4Max|$^*$ & The minimum and maximum allowed 
                    energy of the \\
                  & scattered electron. \\
  \verb|OutputFile|$^*$ & The name of the file in which the generation \\
                  & information is written. \\
  \verb|ICalc|$^*$ & Determines which calculational method to use: \\
                  & =1 : Use the matrix element in terms of invariants (I); \\
                  & =2 : Use the WvdW formalism in terms of \\
                  & \hspace{.7cm} inner products (II); \\
                  & =3 : Use the WvdW formalism in terms of traces of \\
                  & \hspace{.7cm} matrix products(III); \\
                  & =4 : Use I and II; \\
                  & =5 : Use I and III; \\
                  & =6 : Use II and III; \\
                  & =7 : Use all methods.\\
  \verb|IDens|$^*$ & Switch variable (0,1). When set to 1 the density matrix \\
                  & is calculated. \\
  \verb|ICut|$^*$ & Switch variable (0,1). When set to 1 checks for cuts on \\
                  & non-generated variables are made. Note that this switch \\
                  & variable is {\it not} automatically set. \\
  \verb|IBoost|$^*$ & Switch variable (0,1). When set to 1 the resonance \\
                  & momentum in its rest frame is set exactly to $(M,0,0,0)$\\
                  & and $\phi_R$ and $\theta_R$ in this frame are set to 0. \\
  \verb|It(p)Cut|$^*$ & Switch variable (0,1). When set to 1 the cut on \\
                  & $t$ ($t'$) is taken into account and the minimum allowed\\
                  & scattering angle is adapted. \\
  \verb|t(p)Min|$^*$,\verb|t(p)Max|$^*$ & The minimum and maximum allowed 
                    values of \\
                  & $t$ and $t'$. \\
  \verb|IBreitWigner|$^*$ & Switch variable (0,1). When set to 0 the invariant
                    mass
                    \\ &
                    of the $\gamma\gamma$-system, $W$, is set to $M$, the 
                    mass of the generated \\ & resonance. 
                    When set to 1, $W$ is \\ & 
                    distributed according to a Breit-Wigner shape \\ &
                    \\ &
                    $
                      \frac{1}{\pi} \frac{\Gamma}{(W-M)^2+\Gamma^2},
                    $ \\ & \\ &
                    where $\Gamma$ is the total decay width of the resonance.
\end{tabular}

\begin{tabular}{ll}
	            \hspace{3.0cm} &
                    As the photons have a Bremsstrahlung-like character,
                    \\ & 
                    a low $W$ is preferred (high weight). In order to
                    \\ &
                    keep the possibility to generate unweighted events, 
                    \\ &
                    one can choose to generate the invariant 
                    \\ &
                    masses only in the interval $[M_R-n\Gamma,M_R+n\Gamma]$. 
\end{tabular} 

\begin{verbatim}
  MODULE PhysCon
\end{verbatim}

Some physical constants are stored in \verb|PhysCon|. Between brackets
their values are given (All masses and energies in $\mbox{GeV}$):

\begin{tabular}{ll}
  \verb|me| & $m_e$ $(=0.00051099906 \ \mbox{GeV})$; \\
  \verb|Alpha| & $\alpha_{e.m.}$ ($=\frac{1}{137.036}$); \\
  \verb|Pi| & $\pi$ $(=3.14159)$; \\ 
  \verb|BarnConv| & A conversion factor to go from $\mbox{GeV}^{-2}$ to 
  $\mbox{pb}$ \\
            & $(=3.89385 \times 10^5 \ \mbox{pbGeV}^2)$.
\end{tabular}

\begin{verbatim}
  MODULE Resonance
\end{verbatim}
 
Contains some properties of the resonance.

\begin{tabular}{ll}
  \verb|IWav|$^*$ & Determines the wave function of the resonance: \\
                  & =1 : $^1S_0$; =2 : $^3P_0$; =3 : $^3P_1$; =4 : $^3P_2$;
                    =5 : $^1D_2$; \\
  \verb|IRes|$^*$ & Determines the composition of the resonance: \\
                  & =1 : $I=0$; =2 : $I=1$; =3 : $I=1'$; =4 : $c\bar{c}$; 
                    =5 : $b\bar{b}$; \\
  \verb|RName| & Contains the name of the resonance; \\
  \verb|RMass| & Contains the mass of the resonance; \\
  \verb|RWf2|  & Contains the square (of the derivative) of the radial part \\
               & of the wave function in the origin(see \cite{SchuBerGul}); \\ 
  \verb|eq2|   & Contains the value for $e_q^2$ (see \cite{SchuBerGul}); \\
  \verb|GGWidth| & Contains the value of $\Gamma_{\gamma\gamma}$, the 
                   two-photon width; \\
  \verb|Width| & Contains the value of $\Gamma$, the total decay width of the
                 resonance. 
\end{tabular}

\subsection{COMMON blocks}

Some important COMMON blocks:

\begin{verbatim}
  COMMON FourVecs
\end{verbatim}

Contains the four-momenta of the generated particles in the lab-frame.

\begin{tabular}{ll}
  PV1 & $p_1$: four-momentum of incoming positron; \\
  PV2 & $p_2$: four-momentum of incoming electron; \\
  PV3 & $p_3$: four-momentum of outgoing positron; \\
  PV4 & $p_4$: four-momentum of outgoing electron; \\
  PVR & $P_R$: four-momentum of outgoing resonance. 
\end{tabular}

\begin{verbatim} 
  COMMON BFourVecs
\end{verbatim}

Contains the four-momenta in the rest frame of the resonance (RRF) where the
positive $z$-axis is given by the boost direction $\hat{\beta}$. This COMMON
block also contains the five four-momenta of the external particles.

\begin{verbatim}
  COMMON WvdWSpinors
\end{verbatim}

Contains the WvdW spinors of the external particles.

\subsection{Subroutines}

A collection of some important subroutines:

\begin{tabular}{ll}
  \verb|InitPhysCon| & Initializes the physical constants;\\
  \verb|ResProp| & Initializes the resonance properties;\\
  \verb|MC| & Generates an event using scheme I;\\
  \verb|MC2| & Generates an event using scheme II;\\
  \verb|Boost| & Performs the boost to the RRF;\\
  \verb|StoreVar| & Stores the data of the generated event;\\
  \verb|Finish| & Calculates the cross sections after the generation loop \\ 
       & and prints the results.
\end{tabular}

\subsection{Functions}

A collection of some important functions:

\begin{tabular}{ll}
  \verb|RNF100| & Generates a random number uniformly distributed in the \\
  &
  interval $[0,1]$; \\
  \verb|MatrixElement2| & Calculates $\sum |{\cal M}|^2$ using the 
  expression in terms of invariants; \\
  \verb|SpinorMatrixElement2| & Calculates the matrix element squared (and
  the density matrix) \\ & using the WvdW calculation with spinor 
  inner products; \\
  \verb|FastSpinMat2| & Calculates the matrix element squared (and the density
  matrix) \\ & using the WvdW calculation with traces of matrix products \\
  & (preferred method for density matrix calculations); \\
  \verb|Norm| & Calculates the length of the spatial part of a four-vector.
\end{tabular}


\section{Comparison to other MC generators}
The results of the cross section calculations have been checked with the 
results of Galuga \cite{Galuga2}. In this MC generator the same model 
is implemented in the Budnev \cite{Budnev} formalism. Within the errors
we found complete agreement. Density matrix calculations could not be 
checked with any Monte Carlo generator, since only GaGaRes is able to 
calculate them at present. Only by extending the Galuga generator ourselves
checks could be performed and gave agreement.



\section{Conclusions}
GaGaRes is a MC generator which is well-suited for the description of 
resonance production in two-photon physics. It offers several schemes and
calculational methods and 
can be used for every required topology of the outgoing
particles. The program is also able to generate density matrices which form
an essential tool in the description of the decay of the resonances.


\section*{Acknowledgements}
We would like to thank Gerhard Schuler for useful discussions 
and for offering his program Galuga 
to perform the necessary checks.


\appendix

\section{Expressions for $\sum |{\cal M}|^2$}
\label{MatExs}
In the expressions for the total matrix elements squared we have used
the following variables to compactify the expressions
\begin{equation}
  \begin{array}{lcl}
     \Sigma_{n\pm} & = & s^n + s'^n \pm u^n \pm u'^n, \\
     \sigma_{2\pm} & = & ss' \pm uu', \\
     s_{n\pm}      & = & s^n \pm s'^n, \\
     u_{n\pm}      & = & u^n \pm u'^n,\\
     t_{n\pm}      & = & t^n \pm t'^n, \\
  \end{array}
\end{equation}
with the convention that for $n=1$ the $n$ is not written.

This leads us to the following expressions:

\noindent
\renewcommand{\arraystretch}{2.0}
\noindent
\underline{$^1S_0$ amplitude:}
\begin{equation}
  \begin{array}{ll}
    \frac{1}{4} \sum | {\cal M} (^1S_0) |^2   = &
    \frac{c_1^2}{t^2t'^2}
    \left\{ \frac{1}{2} \left[ tt'(s_+^2+u_+^2) 
    -(tt'+\sigma_{2-})^2-(tt'-\sigma_{2-})^2 \right] \right. 
    \\ & 
    + m_e^2 \left[ t_+(s_-^2+u_-^2)+4\sigma_{2-} \Sigma_-
    -4tt'\Sigma_+ + 2s_-u_-t_- \right] 
    \\ &
    +16m_e^4s_+u_+ 
    \left. -64 m_e^6 \Sigma_+ + 256 m_e^8 \right\}.
  \end{array}
  \label{For1S0ExpExp}
\end{equation}

\vspace{1.0cm}

\noindent
\underline{$^3P_0$ amplitude:}
\begin{equation}
  \begin{array}{ll}
    \frac{1}{4} \sum | {\cal M} (^3P_0) |^2  = & \frac{c_2^2}
    {t^2t'^2} \left\{ \frac{B^2}{2} (2\sigma_{2-}^2+tt'
    (s_-^2+u_-^2+2tt'))+Btt'(-s_+ (\sigma_{2-}+tt') \right. \\
    & +u_+(\sigma_{2-}-tt'))+2t^2t'^2\sigma_{2+} + m_e^2 [
    B^2(t_+\Sigma_+^2 - 4\sigma_{2-}\Sigma_-)
    \\ &
    +2tt'B(\Sigma_-^2-2t_+\Sigma_+ 
    + 4tt') - 4t^2t'^2(\Sigma_+ - t_+)] 
    \\ &
    + m_e^4 [4B^2(\Sigma_-^2
    -4t_+\Sigma_+)+32tt't_+B+16t^2t'^2] 
    \left. + 64 m_e^6 B^2 t_+ \right\}.
  \end{array}
  \label{For3P0ExpApp}
\end{equation}
Here $B$ is given by
\begin{equation}
  B=k_1 \cdot k_2 + W^2 = \frac{3}{2} (s+s'+u+u')+t+t'-12m_e^2
  = \frac{3}{2} \Sigma_+ + t_+ - 12 m_e^2.
\end{equation}

\vspace{1.0cm}

\noindent
\underline{$^3P_1$ amplitude:}

\begin{equation}
  \begin{array}{ll}
    \frac{1}{4} \sum | {\cal M}(^3P_1) |^2 = &\frac{c_3^2}
    {W^2 t^2 t'^2}
    \left\{ tt' \left[ tt't_+^2-tt't_+\Sigma_+ +\frac{1}{2}
    t_{2+}\Sigma_{2+} -2tt'\Sigma_{2+} 
    \right. \right. 
    \\ &
    +(t_{2+}-4tt')\sigma_{2+}+\sigma_{2-}\Sigma_{2-}
    -4tt's_+u_++t_+\sigma_{2+}\Sigma_+
    \\ &
    +2t(su'^2+s^2u'+s'u^2+s'^2u) 
    +2t'(su^2+s^2u+s'u'^2+s'^2u')
    \\ & 
    \left.
    +2t^2(su'+s'u)+2t'^2(su+s'u') \right]
    -\sigma_{2-}^2t_{2+} 
  \end{array}
\end{equation}
\[
  \begin{array}{ll}
    \hspace{2.cm} &
    \\ &
    +m_e^2 \left[ t_{3+}\Sigma_+^2-7tt't_+\Sigma_{2+}-26tt'(
    t(su'+s'u)+t'(su+s'u')) 
    \right. 
    \\ &
    +8t_{2+}(ss's_++uu'u_+)+4(t_{2+}+tt')(ss'u_++uu's_+)
    \\ &
    -22tt't_+
    \sigma_{2+} 
    -12tt't_{2+}\Sigma_++32t^2t'^2\Sigma_+-2tt'\Sigma_{3+}
    \\ &
    -6tt'(ss's_+ +uu'u_+)+16t^2t'^2t_+ 
    +2tt'(s_+u_{2+}+u_+s_{2+})
    \\&
    +4t^2(su'^2+s^2u'+s'u^2+s'^2u)
    +4t'^2(su^2+s^2u+s'u'^2+s'^2u') 
    \\ &
    \left.
    -10tt'(stu+st'u'+s't'u+s'tu') \right]
    \\ &
    +16m_e^4\left[ -t_+^2s_+u_++6tt't_+\Sigma_+-t_{3+}\Sigma_+-
    t_{2+}(\Sigma_{2+}+4\sigma_{2+})
    \right. 
    \\ &
    \left.
    + 4tt't_-^2 
    -2(t^2(su'+s'u)+t'^2(su+s'u')) \right]   
    \\ &
    +64m_e^6 \left[ 2(t_{2+}+tt')\Sigma_+ +t_{3+} -5tt't_+ \right]  
    \\ &
    \left.
    -256m_e^8t_+^2.
    \right\}
  \end{array}
  \label{For3P1ExpApp}
\]

\noindent
\underline{$^3P_2$ amplitude:}

\[
  \begin{array}{ll}
    \frac{1}{4} \sum |{\cal M}(^3P_2)|^2 = &
    \frac{c_4^2}{12} \left[ \frac{1}{tt'} \left\{
    \frac{m_e^2}{W^4}[-8\Sigma_+^3-32\sigma_{2+}\Sigma_+
    +64s_+u_+\Sigma_+  \right. \right. 
    \\ &
    +192(ss'u_++uu's_+)]
    +\frac{m_e^2}{W^2}[-4\Sigma_+^2+128\sigma_{2+}-16s_+u_+]
    \\ &
    +m_e^2[-24\Sigma_+]
    +\frac{m_e^4}{W^4}[-192\Sigma_+^2-
    256\sigma_{2+}-640s_+u_+] 
    \\ & 
    +\frac{m_e^4}{W^2}[-128\Sigma_+]
    +m_e^4[96]+\frac{m_e^6}{W^4}
    [3072\Sigma_+]+\frac{m_e^6}{W^2}[512] +\frac{m_e^8}{W^4}
    [-8192]
    \\ &
    +\frac{1}{W^4}[\Sigma_+^4-2s_+u_+(s_+^2+u_+^2)
    +16\sigma_{2-}^2-8s_+u_+\sigma_{2+}
    \\ &
    -16(ss'u_{2+}
    uu's_{2+})
    -4s_{2+}u_{2+}-48ss'uu'] +\frac{1}{W^2}[4\Sigma_+^3
    -8s_+u_+\Sigma_+
    \\ &
    \left.
    -26\sigma_{2+}\Sigma_++20(ss'u_++uu's_+)]+[s_+^2+u_+^2
    +8\sigma_{2+}] \right\}
    \\ &
    +\left( \frac{1}{tt'^2}+\frac{1}{t^2t'} \right)
    \left\{ \frac{m_e^2}{W^4} [ 2 \Sigma_+^4-128\sigma_{2-}^2
    -32\sigma_{2-}\Sigma_{2-}]
    \right.
    \\ &
    +\frac{m_e^2}{W^2}[8\Sigma_+^3
    +8\sigma_{2-}\Sigma_-]+m_e^2[2\Sigma_+^2] 
    +\frac{m_e^4}{W^4}[-32\Sigma_+^3-128s_+u_+\Sigma_+
    \\ &
    +256\sigma_{2-}\Sigma_-]+\frac{m_e^4}{W^2}[-200\Sigma_+^2
    +32s_+u_+]+m_e^4[64\Sigma_+]
    \\ &
    + \frac{m_e^6}{W^4}[512\Sigma_+^2+1024s_+u_+]+\frac{m_e^6}
    {W^2}[1536\Sigma_+]+m_e^6[-256]
    \\ &
    +\frac{m_e^8}{W^4}[-4096\Sigma_+]
    +\frac{m_e^8}{W^2}[-4096]+\frac{m_e^{10}}{W^4}[8192]
  \end{array}
\]
\begin{equation}
  \begin{array}{ll}
    \hspace{2.cm} &
    \left.
    +\frac{1}{W^4}[8\sigma_{2-}^2\Sigma_+]+\frac{1}{W^2}
    [-2\sigma_{2-}^2]
    \right\} 
    +\frac{m_e^2}{tt'^2}[-24(s+u)(s'+u')]
    \\ &
    +\frac{m_e^2}{t^2t'}[-24(s+u')(s'+u)]
    +\frac{1}{t} \left\{ \frac{m_e^2}{W^4}[8\Sigma_+^2
    +64(s+u')(s'+u) \right.
    \\ &
    +128(su'+s'u)
    +32t'\Sigma_+] + \frac{m_e^2}
    {W^2}[-64\Sigma_+-48t'] + m_e^2[48]
    \\ &
    +\frac{m_e^4}{W^4}[-640\Sigma_+-256t']+\frac{m_e^4}{W^2}
    [416]+\frac{m_e^6}{W^4}[2560] +\frac{1}{W^4}[2\Sigma_+^3
    \\ &
    +2t'\Sigma_+^2-8t'(s+u)(s'+u')
    -24(ss'u_++uu's_+)-8(su'(s+u')
    \\ &
    \left.
    +s'u(s'+u))]+\frac{1}{W^2}
    [4\Sigma_+^2-4(s+u)(s'+u')-12\sigma_{2+}] 
    \right\}
    \\ &
    +\frac{1}{t'} \left\{ \frac{m_e^2}{W^4}[8\Sigma_+^2
    +64(s+u)(s'+u')+128(su+s'u')+32t\Sigma_+]
    \right.
    \\ &
    + \frac{m_e^2}
    {W^2}[-64\Sigma_+-48t]+m_e^2[48]
    \\ &
    +\frac{m_e^4}{W^4}[-640\Sigma_+-256t]+\frac{m_e^4}{W^2}
    [416]+\frac{m_e^6}{W^4}[2560] +\frac{1}{W^4}[2\Sigma_+^3
    \\ &
    +2t\Sigma_+^2-8t(s+u')(s'+u)
    \\ &
    -24(ss'u_++uu's_+)-8(su(s+u)+s'u'(s'+u'))]
    \\ &
    \left.
    +\frac{1}{W^2}
    [4\Sigma_+^2-4(s+u')(s'+u)-12\sigma_{2+}] 
    \right\}
    \\ &
    +\frac{1}{t^2t'^2} \left\{ \frac{m_e^2}{W^4}[-56\sigma_{2-}^2
    \Sigma_+-8\sigma_{2-}\Sigma_{3-}-8\sigma_{2-}(s_+uu'-u_+
    ss') 
    \right.
    \\&
    +8\sigma_{2-}(s_+u_{2+}-u_+s_{2+})]
    +\frac{m_e^2}{W^2}[-32\sigma_{2-}^2-8\sigma_{2-}\Sigma_{2-}]
    +m_e^2[16\sigma_{2-}\Sigma_-]
    \\ &
    +\frac{m_e^4}{W^4}
    [416\sigma_{2-}^2+160\sigma_{2-}\Sigma_{2-} +
    8\Sigma_{2+}^2-32s_{2+}u_{2+}] 
    +\frac{m_e^4}{W^2}[8\Sigma_+^3+\sigma_{2-}\Sigma_-
    \\ &
    -32(s_+u_{2+}+u_+s_{2+})-64(ss'u_++uu's_+)]
    +m_e^4[8\Sigma_+^2+64s_+u_+] 
    \\ &
    +\frac{m_e^6}{W^4}
    [-128\Sigma_+^3-512\sigma_{2-}\Sigma_-+512s_+u_+\Sigma_+]
    +\frac{m_e^6}{W^2}[-64\Sigma_-^2]
    \\ &
    +m_e^6[-384\Sigma_+]
    +\frac{m_e^8}{W^4}[512\Sigma_-^2]+m_e^8[1536]+\frac{1}{W^4}
    [2\sigma_{2-}^2\Sigma_{2+}+4\sigma_{2-}^2\sigma_{2+}
    \\ &
    +4\sigma_{2-}^2s_+u_+]
    \left.
    +\frac{1}{W^2}[2\sigma_{2-}^2
    \Sigma_+]+[-4\sigma_{2-}^2] \right\}
    \\ &
    +\left( \frac{1}{t^2}+\frac{1}{t'^2} \right)
    \left\{ \frac{m_e^2}{W^4}[8\Sigma_+^3-32\sigma_{2-}
    \Sigma_-]+\frac{m_e^2}{W^2}[4\Sigma_+^2]+\frac{m_e^4}
    {W^4}[-160\Sigma_+^2
    \right.
    \\ &
    -128s_+u_+]+\frac{m_e^4}{W^2}[-160\Sigma_+]
    +\frac{m_e^6}{W^4}[1536\Sigma_+]+\frac{m_e^6}{W^2}[640]
    +\frac{m_e^8}{W^4}[-4096]
    \\ &
    \left.
    +\frac{1}{W^4}[8\sigma_{2-}^2] 
    \right\}
    +\frac{1}{t^2} \left\{ \frac{m_e^2}{W^4}[8t'\Sigma_+^2]
    +\frac{m_e^2}{W^2}[24(s+u')(s'+u)]
    \right.
    \left.
    +\frac{m_e^4}{W^4}[-128t'
    \Sigma_+] \right.
    \\ &
    \left.
    +\frac{m_e^6}{W^4}[512t'] \right\} 
    +\frac{1}{t'^2} \left\{ \frac{m_e^2}{W^4}[8t\Sigma_+^2]
    +\frac{m_e^2}{W^2}[24(s+u)(s'+u')]+\frac{m_e^4}{W^4}[-128t
    \Sigma_+]
    \right.
  \end{array}
  \label{For3P2ExpApp}
\end{equation}
\[
  \begin{array}{ll}
    \hspace{2.cm} &
    \left. \left.
    \left.
    +\frac{m_e^6}{W^4}[512t] \right\} 
    + \left\{ \frac{m_e^2}{W^4}[64\Sigma_++32t_+]
    +\frac{m_e^2}{W^2}[-144]+\frac{m_e^4}{W^4}[-512]
    +\frac{1}{W^4}[2\Sigma_-^2]
    \right. \right. \right.
    \\ &
    \left. \left.
    +\frac{1}{W^2}[12\Sigma_+
    +6t_+] + [8] \right\} \right].
  \end{array}
\]

\vspace{1.0cm}

\noindent
\underline{$^1D_2$ amplitude:}
\begin{equation}
  \frac{1}{4} \sum | {\cal M} (^1D_2) |^2 = \hat{c}^2
  {\cal F} \frac{1}{4} \sum | {\cal M} (^1S_0) |^2,
\end{equation}
with 
\begin{equation}
  \hat{c}=\frac{c_5}{c_1},
\end{equation}
and
\begin{equation}
    {\cal F} = \frac{1}{2}\left(-t+\frac{(W^2+t_-)^2}{4W^2}\right)
    \left(-t'+\frac{(W^2-t_-)^2}{4W^2}\right)
    +\frac{1}{6}\left(\frac{W^2-t_+}{2}-\frac{W^4-(t_-)^2}{4W^2}\right)^2.
\end{equation}

Note that for the $^1D_2$ case the overall factor $\frac{e^4}{t^2t'^2}$
is included in the $^1S_0$ factor.    

\vspace{1.0cm}


\begin{thebibliography}{99}

  \bibitem{SchuBerGul} G.\ A.\ Schuler, F.\ A.\ Berends, R.\ van Gulik, Nucl.\ 
                  Phys.\ {\bf B523} (1998) 423.

  \bibitem{Linde} F.\ L.\ Linde, Thesis, Leiden University (1988).

  \bibitem{Buijs} A.\ Buijs, W.\ G.\ J.\ Langeveld, M.\ H.\ Lehto, D.\ J.\ 
                  Miller, Comp.\ Phys.\ Com.\ {\bf 79} (1994) 523.

  \bibitem{Dittmaier} S.\ Dittmaier, Phys.\ Rev.\ {\bf D59} (1999) 016007. 

  \bibitem{Budnev} V.\ M.\ Budnev et al., Phys.\ Rep.\ {\bf 15C} (
                   1975),181.
 
  \bibitem{Vermaseren} J.\ A.\ M.\ Vermaseren, FORM2, Computer Algebra
                       Nederland.

  \bibitem{BerendsGiele} F.\ A.\ Berends, W.\ T.\ Giele, Nucl.\ Phys.\ 
                         {\bf B294} (1987) 700; \\ 
                         W.\ T.\ Giele, Thesis, Leiden University (1989).



  \bibitem{Pirani} 
                   H.\ Kuijf, Thesis, Leiden University (1991). 

  \bibitem{Gulik} R.\ van Gulik, Nucl. Phys. Proc. Suppl {\bf 82} (2000) 311.

  \bibitem{DavBerKlei} F.\ A.\ Berends, P.\ H.\ Daverveldt, R.\ Kleiss, 
                   Comp.\ Phys.\ Com.\ {\bf 40} (1986) 271.

  \bibitem{Galuga2} G.\ A.\ Schuler, Comp.\ Phys.\ Com.\ 
                    {\bf 108} (1998),279.

\end{thebibliography}
\end{document}